\documentclass[11pt,english,twoside]{article}
\pdfoutput=1
\usepackage{jheppub}
\usepackage[numbers]{natbib}
\usepackage{lscape}
\usepackage{amsmath}
\usepackage{xcolor}
\usepackage{graphicx}
\usepackage{dcolumn}
\usepackage{bm}
\usepackage{tikz}
\usetikzlibrary {arrows.meta} 
\usetikzlibrary{positioning}

\def\-{\hphantom{-}}

\def\s2{\frac{1}{\sqrt2}}

\def\beq{\begin{equation}}
\def\eeq{\end{equation}}
\def\bea{\begin{align}}
\def\eea{\end{align}}
\def\beqa{\begin{eqnarray}}
\def\eeqa{\end{eqnarray}}

\def\Z{{\mathbb Z}}

\def\mg{m_{3/2}}
\def\mg2{m^2_{3/2}}

\def\Dsl{\,\raise.15ex\hbox{/}\mkern-13.5mu D} 

\def\f{\text{flux}}
\def\H{\text{H}}

\newcommand{\eq}[1]{\begin{equation}
                     \begin{split} #1 \end{split}
                     \end{equation}}

\begin{document}

\title{Some remarks on Swampland conjectures, fluxes and K-theory in IIB toroidal compactifications}

\author[b]{Cesar Damian,} 
\author[a]{Oscar Loaiza-Brito}

\affiliation[a]{Departamento de F\'isica, Universidad de Guanajuato, \\
Loma del Bosque No. 103 Col. Lomas del Campestre C.P 37150 Leon, Guanajuato, Mexico.}
\affiliation[b]{Departamento de Ingenier\'ia Mec\'anica, Universidad de Guanajuato, \\
Carretera Salamanca-Valle de Santiago Km 3.5+1.8 Comunidad de Palo Blanco, Salamanca, Mexico}

\emailAdd{cesaredas@fisica.ugto.mx}
\emailAdd{oloaiza@fisica.ugto.mx}

\date{\today}

\abstract{ In this note, we present a scenario in which the existence of dS vacua is jeopardized by topological transitions between non-BPS branes and fluxes in type IIB toroidal compactifications. We focus our study on a six-dimensional torus compactification modded out by an orientifold three-plane. In particular, we consider the presence of non-BPS five branes, which allows the construction of apparent stable dS vacua. Since moduli stabilization requires the presence of NS-NS fluxes, we also describe the minimal required conditions to avoid the appearance of Freed-Witten anomalies. Cancellation of these anomalies implies a topological transformation between non-BPS states and fluxes with the same discrete K-theory charge. After the transition, we observe that the scalar potential undergoes major changes, implying that the dS vacua are unstable. We discuss a scenario in which, even if the discrete K-theory charge is globally cancelled, the above transitions could still be relevant, allowing the existence of short-lived dS minima. We briefly comment on how these implications seem to be in concordance with the Refined de Sitter and Trans-Planckian Censorship Conjectures.
}
\arxivnumber{}


\maketitle


\section{Introduction}

The Swampland project has received considerable attention in recent years. It offers an opportunity to examine the relevant characteristics of effective theories that can be completed at high energies to a quantum theory of gravity. (For reviews, see \cite{Palti:2019pca, vanBeest:2021lhn, Grana:2021zvf, Agmon:2022thq}.) One of the project's main issues concerns a deeper understanding of the Refined de Sitter Conjecture (RdSC) \cite{Garg:2018reu, Ooguri:2018wrx}, which states that a stable de Sitter (dS) vacuum belongs to the Swampland at least in asymptotic regions of the moduli space, while a short-lived dS could be allowed in the context of the Trans-Planckian Censorship Conjecture (TCC) \cite{Bedroya:2019snp}.\\

Studying different scenarios and searching for stable dS vacua is a way to test the validity of the above proposals. Specifically, we are interested in the role played by K-theory \cite{Witten:2000cn} and torsional cohomology in phenomenologically viable scenarios \cite{Marchesano:2006ns, Camara:2011jg, BerasaluceGonzalez:2012zn, Garcia-Compean:2013sla, Marchesano:2014mla} as well as possible implications on some of the Swampland conjectures.\\

Our interest lies in two important features that arise from K-theory when applied in string theory. Firstly, it classifies objects that remain invisible to integer cohomology, such as the so-called non-BPS branes, constructed by a pair of brane and anti-brane in the presence of orientifolds \cite{Frau:1999qs}. Incorporating such states has led to important consequences in string phenomenology \cite{Marchesano:2004yq, Marchesano:2004xz}. Secondly, recent studies have shown that the presence of discrete K-theory charge can compromise the RdSC for a non-zero total K-theory charge over the compact internal manifold \cite{Blumenhagen:2019kqm}. However, since a theory with a non-zero (integer or discrete) K-theory tadpole is not quantum mechanically consistent \cite{Uranga:2000xp, Garcia-Etxebarria:2018ajm, GarciaEtxebarria:2019caf}, a relationship between the cancellation of discrete K-theory charge and trivial cobordism class has been recently proposed \cite{McNamara:2019rup, Blumenhagen:2021nmi, Andriot:2022mri}.\\

A scenario where non-BPS states are considered as a possibility to construct a dS vacuum in type IIA CY compactifications in the presence of fluxes is studied in \cite{Blumenhagen:2019kqm} (a similar approach in a Type IIA context is considered in \cite{Marchesano:2014mla}). Within the many possibilities, it is mentioned that a non-BPS $\widehat{D6}$-brane is discarded since it is potentially anomalous under the Freed-Witten (FW) anomaly \cite{Freed:1999vc}. This is because it is wrapped on a 3-cycle in the internal manifold which can support the NS-NS 3-form flux $H_3$. However, under a T-dual perspective (on a coordinate where the NS-NS flux is not supported), such a non-BPS state transforms into a $\widehat{D5}$-brane wrapping a 2-cycle in the internal manifold. Since a NS-NS flux cannot be supported in the same cycle, it seems that the brane is free of FW anomalies.\\

Based on the above, we want to discuss the implications of incorporating non-BPS states in toroidal IIB compactifications, with the purpose of finding possible stable dS vacua and searching for the T-dual manifestation of the FW anomaly in this scenario. To simplify the discussion, we focus on an isotropic $T^6$ torus with three moduli: the complex structure $U$, the complex axio-dilaton $S$, and the K\"ahler modulus $T$. We also consider an example involving a Calabi-Yau compactification with no complex structure, which emulates a toroidal compactification with a fixed complex structure.\\

In the toroidal case, we find that 
the non-BPS $\widehat{D5}$ seems to be essential to have a SUSY AdS minimum that can be uplifted to a short-lived dS minimum, placing this scenario in the Landscape. In the CY case without complex structure, we find that a scalar potential with a runaway direction along K\"ahler moduli, in which non-perturbative effects are considered, can be lifted to an apparently stable dS vacuum by incorporating a non-BPS $\widehat{D5}$-brane.  The total scalar potential is also very flat, as in the toroidal case, and it seems that the presence of a non-BPS state places the scenario in the Landscape.\\

Since the T-dual picture in type IIA, which considers $\widehat{D6}$, is FW anomalous, we expect to have some manifestation of it in these scenarios. We find that indeed, the anomaly is present but manifested by the presence of instantonic branes and topological transformations between branes and fluxes.  In our case, although the non-BPS brane is stable in the sense that it wraps a two-cycle which can not support a three-form flux, it can end on an instantonic 7-brane wrapping a discrete four-cycle which supports the NS-NS flux, inducing a topological transition between branes and fluxes\footnote{The presence of instantonic 7 $(p,q)$-branes was studied in \cite{Loaiza-Brito:2004ajy} to show that the $SL(2,\mathbb{Z})$-symmetry in ${\cal N}=4$ super Yang-Mills theories can be interpreted as equivalence classes in K-theory. Similar results, extended to the context of the cobordism conjecture are aborded in \cite{Dierigl:2022reg, Debray:2023yrs}.}. This transition is manifested as a transformation of the non-BPS state $\widehat{D5}$-brane into a discrete one-form flux $f_1$. As a result of this transition, the scalar potential undergoes a significant change, and the existence of the original vacuum can no longer be guaranteed.\\

This is particularly interesting because, as many studies indicate, the total discrete K-theory charge must vanish (in our case, by having an even number of $\widehat{D5}$). The above transition opens up the possibility of achieving a zero discrete K-theory tadpole through a configuration of non-BPS states and discrete fluxes. Since both of these components have a different dependence on the K\"ahler modulus, the existence of a stable dS vacuum becomes possible. However, under the topological transition between non-BPS branes and discrete fluxes  the scalar potential can undergo major changes, rendering the dS minima unstable.\\

In summary, the incorporation of non-BPS states in toroidal and Calabi-Yau compactifications, both with and without complex structure, can lead to the possibility of constructing stable dS vacua. However, the presence of Freed-Witten anomalies must be carefully considered in these scenarios, as they can manifest in the form of topological transformations between branes and fluxes, ultimately rendering the dS minima unstable.
Additionally, the flatness of the scalar potential, together with the short time of stability implied by the TCC, places these scenarios in the Landscape, highlighting the importance of further investigations into the properties of the Landscape and the role of non-BPS states and topological transitions in dS vacua construction.
Overall, these results offer insights into the complex interplay between topology, non-BPS states, and fluxes in the construction of dS vacua in string theory, paving the way for future developments in this field.\\

 Our work is organized as follows: In Section \ref{sec:2}, we briefly review the construction of non-BPS states by wrapping branes on homology cycles and their relation to K-theory. More importantly for our work, we describe how the presence of NS-NS fluxes, which are sources for FW anomalies, can be incorporated in a homology description of states through the Atiyah-Hirzebruch Spectral Sequence (AHSS). The AHSS tells us which branes are unstable to decay into a topologically equivalent configuration and establishes the existence of a transformation between branes and fluxes, interpreted in terms of the Maldacena-Moore-Seiberg instantons. We also give a brief description of how the AHSS is implemented in the case of discrete components of homology.
In Section \ref{sec:3}, we present two examples in which it is possible to construct an apparent dS (de Sitter) vacuum using a $\widehat{D5}$-brane, which in principle would put such scenarios in the swampland due to the RdSC (de Sitter conjecture in the swampland program). We discuss interesting specifics about the model, as well as possible sources of instabilities in the construction of the vacuum.
Section 4 is devoted to describing the topological transition between $\widehat{D5}$ and a discrete flux $f_1$ driven by the NS-NS flux in terms of K-theory and the AHSS, which makes the corresponding dS vacua unstable. We propose that the total K-theory charge can be cancelled by the presence of a non-BPS state and a discrete flux while having a scalar potential with a dependence on the moduli, allowing for the construction of an apparent dS vacuum. However, due to these transitions, we expect the minimum to be unstable. We also briefly discuss how these results seem to be in agreement with the RdSC and the TCC. Note that many interesting studies have emerged since the original version of this manuscript, such as \cite{Blumenhagen:2022bvh}.\\

\section{\label{sec:2}Discrete K-theory charge and non-BPS branes}

In this section, we review some important concepts related to the existence of non-BPS branes in compactified models, which can be classified through K-theory. We also discuss the vanishing of the total discrete K-theory and alternative descriptions of these states by wrapping branes on homology cycles. In our models, a NS-NS flux is always present for moduli stabilization, which requires a refinement of (co)homology groups in its presence. This refinement means that we must discard those branes that wrap cycles supporting the NS-NS flux due to the FW anomaly. We must also carefully interpret those branes that end on higher dimensional ones in order to cancel such an anomaly. We briefly overview some important consequences, such as the brane-flux transition.\\

\subsection{Non-BPS states in K-theory}
As shown in \cite{Uranga:2000xp}, the total discrete K-theory charge from non-BPS $\widehat{Dp}$-branes must vanish in any T-dual version of Type I theory on a compact space. For instance, consider type I compactified on a two-dimensional torus $T^2$ with a non-BPS $\widehat{D7}$-brane located at a point. The discrete charge is measured by the K-theory group $\text{KO}(S^2)=\Z_2$. However, the system is not consistent because a $\widehat{D0}$-brane surrounding any point in $T^2$ (also carrying a discrete $\Z_2$ charge measured by $\text{KO}(S^9)=\Z_2$) would acquire a phase. It is then concluded that the total discrete K-theory charge must vanish. Further discussions about more subtle arguments concerning K-theory charge cancellation can be found in \cite{Garcia-Etxebarria:2018ajm, GarciaEtxebarria:2019caf}.\\

After applying T-duality, this constraint remains valid. To see this, consider type I theory compactified on a factorizable $T^6$, with $D7_i$-branes wrapping coordinates transverse to the $i$-th torus $T^2$. The same inconsistency arises from the corresponding $\widehat{D0}$-branes surrounding points in $T_i^2$. Now, let us perform T-duality on all internal coordinates, resulting in type IIB on $T^6$ with sixty-four orientifold three-planes $O3^-$ located at fixed points in $T^6$. The non-BPS $\widehat{D7}$-branes become $\mathbb{Z}_2$-charged $\widehat{D5}$-branes wrapping internal two-cycles. In fact, these non-BPS states can be regarded as a pair of $D5$-$\overline{D5}$ wrapping a two-cycle in the covering space. Since the tachyon arising in the spectrum of strings with one end attached to a $D5$-brane and the other to a $\overline{D5}$-brane is projected out by the orientifold, the system is stable with no integer RR charge. The inconsistency of having a single (or an odd number of) $\widehat{D5}$ is also reflected at the quantum level by the presence of a global gauge anomaly on the four-dimensional field theory at the probe approximation of $\text{D}3$-branes \cite{Uranga:2000xp, LoaizaBrito:2001ux}.\\

A formal calculation of the discrete states spectrum in a given scenario, such as compactification on a $T^6$ in the presence of orientifold three-planes, requires the computation of relevant K-theory groups. Moreover, since we are considering fluxes, these K-theory groups undergo modifications, resulting in what is called twisted K-theory groups \cite{Evslin:2006cj},  which computation  can be rather involved.\\

\subsection{Non-BPS by wrapping branes on (refined) homology cycles}
  
  An alternative way to describe the states mentioned earlier involves wrapping $Dp$-branes on homology cycles, which can be either integer or torsional. This approach has two advantages over the more complicated K-theory counterpart. Firstly, computing homology cycles is generally simpler. Secondly, we can investigate the effects of fluxes by implementing an algorithmic process that refines the integer cohomology through a set of finite steps. The end result is a set of forms that actually belong to the K-theory group. This refinement process is known as the Atiyah-Hirzebruch Spectral Sequence (AHSS), which can be reviewed in Appendix C of \cite{Diaconescu:2000wy} and it has a physical interpretation through the Maldacena-Moore-Seiberg (MSS) instantons, as explained in \cite{Maldacena:2001xj}.\\

Let us provide a simple and general description of the AHSS in terms of branes and fluxes, which are restricted to the internal space $X_6$. We begin by considering stable D$(p+3)$-branes as those branes that wrap non-trivial homology cycles $\Sigma_p\in \H_p(X_6;\mathbb{Z})$. Due to the presence of orientifold planes, the homology cycles can have torsional components, which are denoted by $\text{Tor}\,\H_p(X_6;\mathbb{Z})$. Therefore, a $p$-cycle $\Sigma_p$ is related to a form $\omega_{6-p} \in \H^{6-p}(X_6;\mathbb{Z})$ through Poincar\'e Duality. We then associate a cohomology form $\omega_{6-p}$ with a stable D$(p+3)$-brane.\\

The first refinement under the AHSS is as follows: consider the map
\eq{
d_3: \H^q(X_6;\mathbb{Z})\rightarrow \H^{q+3}(X_6;\mathbb{Z}),
}
such that
\eq{
d_3 (\omega_q)= \omega_q\wedge H_3,\\
}
where $H_3\in \H^3(X_6;\mathbb{Z})$ is the NS-NS 3-form flux\footnote{In case of torsional components, as in our case, the wedge product must be replaced by the cap product.}. In this context, a D$(p+3)$-brane is now stable (meaning free of FW anomalies) if the associated form $\omega_{6-p}$ is closed under $d_3$, i.e., $d_3(\omega_{6-p})=0$, and it is not exact under $d_3$, i.e., it is not of the form $\omega_{6-p}=d_3(\sigma_{3-p})=\sigma_{3-p}\wedge H_3$, for some $\sigma_{3-p}\in\H^{3-p}(X_6;\mathbb{Z})$. The coset group is defined as\\
\eq{
E_3^p(X_6;\mathbb{Z})=\frac{\{\omega_p \in \H^p(X_6;\mathbb{Z})| d_3(\omega_p)=0\}}{\{\omega_p | \omega_p=d_3(\sigma_{p-3}), \sigma_{p-3}\in\H^{p-3}(X_6;\mathbb{Z})\}}\\
}
not only removes all FW anomalous branes but also discards those which are $d_3$-exact. This has a  physical interpretation we proceed to describe \cite{Maldacena:2001xj}. Consider a D$(p+3)$-brane represented by $\omega_{6-p}$. If $\omega_{6-p}=d_3(\sigma_{3-p})$, it means that a D$(p+3)$-brane is unstable if it encounters a brane represented by the form $\sigma_{3-p}$. Due to the dimensionalities, it is not hard to realize that this brane is actually a time-localized D$(p+5)$ brane called instantonic. It is wrapped on an internal $(p+2)$-cycle in $\H_{p+2}(X_6;\mathbb{Z})$ and the 3 external spatial coordinates. Since this brane is localized in time, it leaves after its disappearance (since it is unstable) a magnetic RR flux given by a $(3-p)$-form in $\H^{3-p}(X_6;\mathbb{Z})$.\\

This has a very important interpretation: a D$(p+3)$-brane wrapping a $p$-cycle in $X_6$, under the presence of a NS-NS flux $H_3$ not supported in the same $p$-cycle, transforms into a flux $F_{3-p}$ and the same units of $H_3$. Hence, the coupling $F_{3-p}\wedge H_3 $  carries the same units of the disappeared D$p$-branes. \\

The AHSS can have many subsequent steps through mapping of higher order. However, in string theory, it seems that the only relevant step is the one described above. In this sense, the K-theory group classifying $(p+3)$-dimensional objects is, up to a short exact sequence, the coset group $E_3^p(X_6;\mathbb{Z})$.

\subsection{The non-BPS $\widehat{D5}$-brane}
\begin{figure}[htbp]
   \centering
   \includegraphics[scale=0.3]{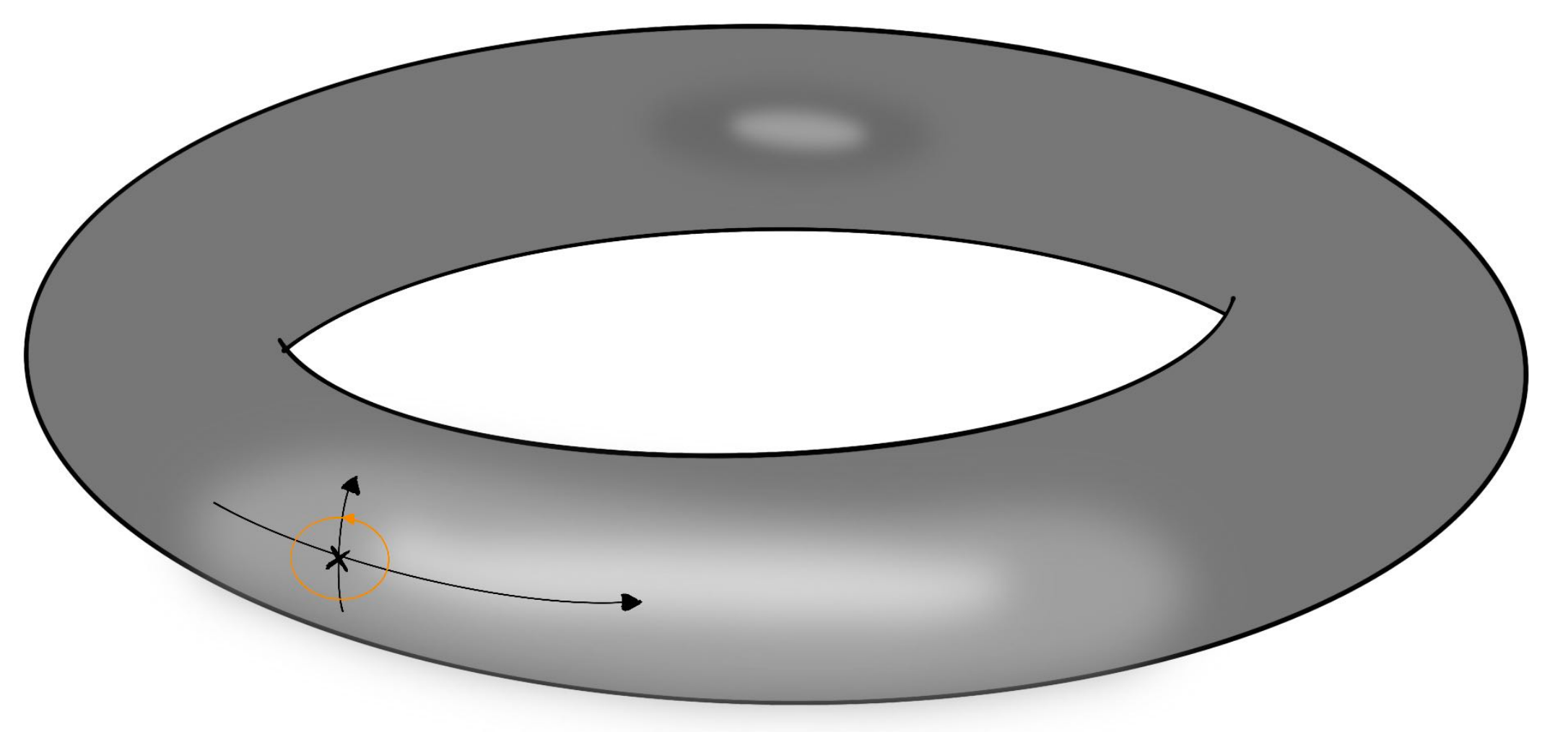} 
   \caption{2-Torus showing the torsional cycle.}
    \begin{picture}(0,0) 
  \put(-25,65){$x_1$}
   \put(-100,90){$x_4$}
   \put(-80,75){$\Z_2$ 1-cycle.}
     \end{picture}
   \label{fig:torus}
\end{figure}

In our case, the six-dimensional torus $T^6$ undergoes a discrete action of $\mathbb{Z}_2$, resulting in the coset space $T^6/\mathbb{Z}_2$. In terms of homology, there exist torsional cycles in $T^6/\mathbb{Z}_2$ that correspond to $\mathbb{Z}_2$ \cite{Frey:2002hf}. As an illustration, let us consider the torus shown in Figure \ref{fig:torus}.
Therefore, consider the 1-cycle in the coset space that encircles the fixed points $x_1 = x_4 = \pi$ corresponding to the first torus $T_1^2$. A brane that wraps twice around this cycle shrinks to a point, as twice the cycle is the boundary of a 2-dimensional submanifold on the covering space $T_1^2$. As a result, the homology group $\H_1(T^6/ \mathbb{Z}_2; \mathbb{Z})$ has torsion, specifically $\text{Tor}\, \H_1(T^6/ \mathbb{Z}_2; \mathbb{Z}) \simeq \text{Tor}\, \H^2(T^6/\mathbb{Z}_2; \mathbb{Z}) = \mathbb{Z}_2$. The same is true for other relevant cycles as\\
\begin{eqnarray}
\rm{Tor}~\rm{H}_2(T^6/\Z_2; \Z)\simeq\rm{Tor}~\rm{H}^3(T^6/\Z_2;\Z)=\Z_2,\nonumber\\
\rm{Tor}~\rm{H}_0(T^6/\Z_2; \Z)\simeq\rm{Tor}~\rm{H}^1(T^6/\Z_2;\Z)=\Z_2.
\end{eqnarray}

It is important to note that the discrete part is obtained by removing the fixed points where the orientifolds are placed\footnote{Strictly speaking, we are considering the space ($T^6-\{0\})/\mathbb{Z}_2$, which results from removing the fixed points under the action of $\mathbb{Z}_2$, because the fixed points do not belong to the torsional part of the(co)homology groups.}. Since we are dealing with a factorizable six-dimensional torus, there are also discrete cycles of higher dimensionality, formed by the product of integer cycles (which originate directly from the covering space) and/or the discrete ones (half-integer modulo 1). Therefore, higher-dimensional homology groups also have torsion components. For instance, consider a type IIB D5-brane wrapped on a discrete two-cycle in $T^6/\mathbb{Z}_2$. This brane is T-dual to the non-BPS $\widehat{D7}$-branes in type IIA  considered in \cite{Blumenhagen:2019kqm}. Note that discrete D5-branes constructed in this manner define non-trivial cohomology classes, in contrast to D7-branes in Type I. \\

This is in general a result that reflects the difference between a (co)homology and a K-theory classification, since there are (co)homology states that are uplifted to trivial or equivalent classes in K-theory  while there are other states in K-theory that are not seen by homology. In order to match both descriptions, we require refining the cohomology by the AHSS and solving some short exact sequences \cite{Bergman:2001rp, LoaizaBrito:2001ux, Loaiza-Brito:2004ajy}. This is due to the fact that T-duality acts not within cohomology but in derived categories \cite{Hori:1999me, Sharpe:1999qz}.\\

Hence, although it is possible to find some differences between a classification of states by wrapping D-branes on homology cycles and the K-theory counterpart, the refinement due to the presence of the NS-NS flux will give us the actual states in K-theory together with some possible relations between different states. \\

In the presence of torsional components in the (co)homology groups, the AHSS involves mappings between them, as stated in \cite{Bergman:2001rp}. In our case, we need to consider the possible mappings between the torsional components of the (co)homology groups related to the non-BPS $\widehat{D5}$-brane and the implications of brane-flux transitions. Before studying these effects, we will demonstrate the relevance of adding a  $\widehat{D5}$-brane in constructing dS extreme in specific scenarios, and then discuss how they align with  the TCC and the RdSC. We shall come back to the implications of the AHSS in Section 4.\\

\section{\label{sec:3} Constructing a dS extreme through non-BPS states}
Let us consider a compactification of type IIB on $T^6$, where an $O3^-$-plane acts on a single non-BPS $\widehat{D5}$ brane. This brane can also be viewed as a $D5$-brane that wraps an internal torsional 2-cycle. It is worth noting that even though this state has no RR charge\footnote{This can be explained by the fact that a non-BPS $\widehat{D5}$-brane is formed by a $D5$-brane and an anti-$D5$-brane. From a homology perspective, this object wraps a 2-cycle in $\mathbb{Z}_2$, which is related to a 3-form, also in $\mathbb{Z}_2$. Therefore, it does not carry any integer RR charge.}, it does carry a topological $\Z_2$ charge, which means it cannot decay into the vacuum. The non-BPS state has a positive contribution to the four-dimensional scalar potential of the effective theory, as mentioned in \cite{Blumenhagen:2019kqm}. In the case of an isotropic factorizable $T^6$, the three complex moduli can be expressed as
\eq{ \label{eq:mod}
U= u + i v \,, \quad S = s- i c \,, \quad T = \tau+ i \rho \,,
}
where $U$ is the complex structure modulus, $s= e^{-\phi}$ is the dilaton and $c$ is the RR zero form. The real part of $T$ is denoted by $\tau$, which can be expressed as $\tau = e^{-\phi} t^2$. Additionally, $\rho$ represents the period of the RR four-form. The perturbative K\"ahler potential is written as
\eq{
K = -3\ln \left( U + \bar U \right)- 3\ln \left( T+ \bar T \right) - \ln \left( S+ \bar S \right) \,.
}
In this note, we shall not consider quantum corrections to $K$. As usual, the moduli $S$ and $U$ are stabilized by turning on appropriate NS-NS and RR 3-form fluxes inducing an effective potential determined by  the perturbative Gukov-Vafa-Witten potential \cite{Gukov:1999ya} 
\eq{ \label{eq:GVW}
{\cal W}_{\text{flux}}=\int G_3\wedge\Omega=P_1(U)-iSP_2(U),
}
 where $G_3=F_3-iS H_3$ and $P_1(U)$ and $P_2(U)$ are cubic polynomials on $U$ with real coefficients fixed by the 3-form fluxes  (for further details, see \cite{Aldazabal:2006up} and the references therein). The K\"ahler modulus can be stabilized by incorporating non-perturbative corrections to the superpotential, which can be achieved by including Euclidean $D3$-branes or, as is the case here, non-BPS $\widehat{D5}$-branes. \\
 
 In the following, we will present some specific scenarios in which we can construct a dS extremal point in the scalar potential by adding the contribution of a non-BPS $\widehat{D5}$-brane and by considering non-perturbative terms in the superpotential. However, as we shall discuss, the inclusion of non-BPS states required to obtain dS extrema tends to flatten the scalar potential, which can lead to instability.\\
 

\subsection{Model 1: No dS without uplifting}
We are interested in establishing the conditions to have a non-SUSY minimum of the scalar potential generated by ${\cal W}_\f$ where the K\"ahler direction be uplifted by the inclusion of a non-BPS $\widehat{D5}$-brane. For that,  consider the case in which the complex structure $U$ is  fixed at $U_0$ due to the conditions $D_UP_1=D_UP_2=0$. This means that the K\"ahler derivatives of both polynomials $P_1$ and $P_2$ with respect to $U$, share a common root while $P_i(U_0)\ne 0$. This condition ensures that  $\partial_U V_{\f} (U_0)= 0$, with
\eq{
V_\f= e^K\left(|D_a W_\f|^2 -3|W_\f|^2\right),
}
where $|D_a{\cal W}_{\f}|^2= (D_a{\cal W}_{\f})( D_{\overline{a}}{\cal \overline{W}}_{\f}) K^{a\overline{a}}$ and $a=U,S,T$. Although the vanishing of $D_UP_i$ establishes a relationship between  RR and NS-NS fluxes, it allows the complex structure to be stabilized independently of $S$ and $T$, while keeping $\mathcal{W}_{\f}\ne 0$. Therefore, by maintaining $D_S{\cal W}_{\f}\ne 0$, the no-scale scalar potential that results from fluxes can be reduced to:
\eq{
V_{\f}=\frac{|D_S{\cal W}_{\f}|^2}{2^7 u_0^3 s \tau^3}.
}
Notice that $|D_S{\cal W}_{\f}|^2$ is a function on $S$ and $\overline{S}$, actually a quadratic polynomial on $s$ over the complex,  given by
\begin{eqnarray}
|D_S{\cal W}_{\f}|^2&=& \frac{(S+\overline{S})}{2} \left(|P_1(U_0)|^2 +|S|^2|P_2(U_0)|^2+2\text{Im}(SP_1(U_0)\overline{P}_2(U_0))\right), \nonumber\\
&=& \alpha s^2+\beta s+\gamma,
\end{eqnarray}
with $\alpha, \beta, \gamma$ being complex numbers depending on $P_1$ and $P_2$ and fixed by $U_0$ and $c$. We shall restrict our description to the case $c=0$ which is indeed a solution.\\

Now, since $V_\f$ is positive definite, the minimum value it can have is zero. This is a minimum in $U$ and $S$ directions but not in $T$.  In order to search for the possibility to stabilize the K\"ahler modulus, let us incorporate the non-BPS brane and consider the more general scalar potential 
\eq{V_{\text{eff}}= V_\f+V_{\widehat{D5}},}
with the contribution $V_{\widehat{D5}}$ given by
\begin{equation}
V_{\widehat{D5}}=\frac{A_{\widehat{D5}}}{s^{1/2}\tau^{5/2}},
\end{equation}
which comes from the dimensional reduction of the Dirac-Born-Infeld action for the non-BPS $\widehat{D5}$-brane \cite{Kachru:2002gs, Blumenhagen:2019kqm}. In this case, the effective scalar potential can be written as
\begin{eqnarray} \label{eq:scalar}
V_{\text{eff}}&=&\frac{|D_S{\cal W}|^2}{2^7 u_0^3 s \tau^3}+\frac{A_{\widehat{D5}}}{s^{1/2}\tau^{5/2}},\nonumber\\
&=&\frac{1}{2^7 u_0^3 \tau^3}\left(\alpha s+\beta +\frac{\gamma}{s}\right)+\frac{A_{\widehat{D5}}}{s^{1/2}\tau^{5/2}},
\end{eqnarray}
where $A_{\widehat{D5}}$ depends on geometric factors of the internal space and does not depend on any moduli of the DBI action. Since $A_{\widehat{D5}}$ is positive (it is constructed from twice the tension of a $D5$-brane) it is possible to simultaneously solve the equations $\partial_s V_{\text{eff}}=0$ and $\partial_\tau V_{\text{eff}}=0$ and fix the axions to their solutions $s=s_0$ and $\tau=\tau_0$, where $s_0$ is a solution of the quadratic equation
\eq{\alpha\left(\frac{2^7\cdot 5}{3} u_0^3\beta-1\right)s^2-\beta s - \frac{8}{3}\gamma = 0,}
while $\tau_0$ is given by
\begin{eqnarray}
\tau_0^{1/2}&=& -\frac{3}{2^{6}\cdot 5}\frac{ |D_S \mathcal{W}(s_0)|^2}{ A_{\widehat{D5}} s_0^{1/2} u_0^3},
\end{eqnarray}
which is physically viable for $A_{\widehat{D5}}$ negative. Since $A_{\widehat{D5}}$ is positive by construction, it is not possible to stabilize the K\"ahler modulus, and extra structure is required.

 
\subsection{Model 2: dS by uplifting an AdS}
In this example, $-$contrary to the previous case$-$ we succeed in constructing a dS extreme  by uplifting an AdS critical point (similarly as in a KKLT scenario \cite{Kachru:2003aw})  by adding non-perturbative corrections to the superpotential $\mathcal{W}$ together with the inclusion of a non-BPS state,  which plays the role of the anti $D3$-branes in KKLT scenarios.\\

As a first step, let us consider the construction of a stable AdS vacuum in the context of KKLT, and study whether the inclusion of a non-BPS state $\widehat{D5}$ would uplift the vacuum to a dS one. To do so, consider both, the superpotential  $\mathcal{W}_{\text{flux}}$ in Eq. (\ref{eq:GVW}) and $\mathcal{W}_{\text{inst}}$ induced by Euclidean D3 branes wrapping internal 4-cycles, given by
 \eq{
 \mathcal{W}_{\text{inst}} = A (U) \exp \left( -\lambda T \right)
 \label{inst}
 }
where $A(U)$ is a complex structure-dependent one-loop determinant. As in the previous case, the complex structure is fixed by the condition $D_U \mathcal{W} = 0$, and in consequence, the pre-factor acquires a constant value $A(U_0)$ while the axio-dilaton is fixed by $D_S\mathcal{W}_{\text{flux}}=0$. The complete superpotential 
\eq{
\mathcal{W} = \mathcal{W}_{\text{flux}} + \mathcal{W}_{\text{inst}}
}
breaks the no-scale structure and allows to stabilize the K\"ahler modulus by $D_T\mathcal{W}=0$. The scalar potential has a supersymmetric minimum at
\eq{
V_{\text{AdS}} = - 3 e^{K} |\mathcal{W}_0|^2 
}
which fixes the K\"ahler modulus at large values in an AdS minimum when $\mathcal{W} (U_0,S_0)=\mathcal{W}_0 \ll 1$ (for a Calabi-Yau compactification this can be realized by an appropriate flux configuration and the use of the racetrack mechanism \cite{Demirtas:2019sip}). Notice that this step is compatible with the non-SUSY AdS conjecture \cite{Ooguri:2016pdq}.\\

\begin{figure}[htbp]
   \centering
   \includegraphics[scale=0.5]{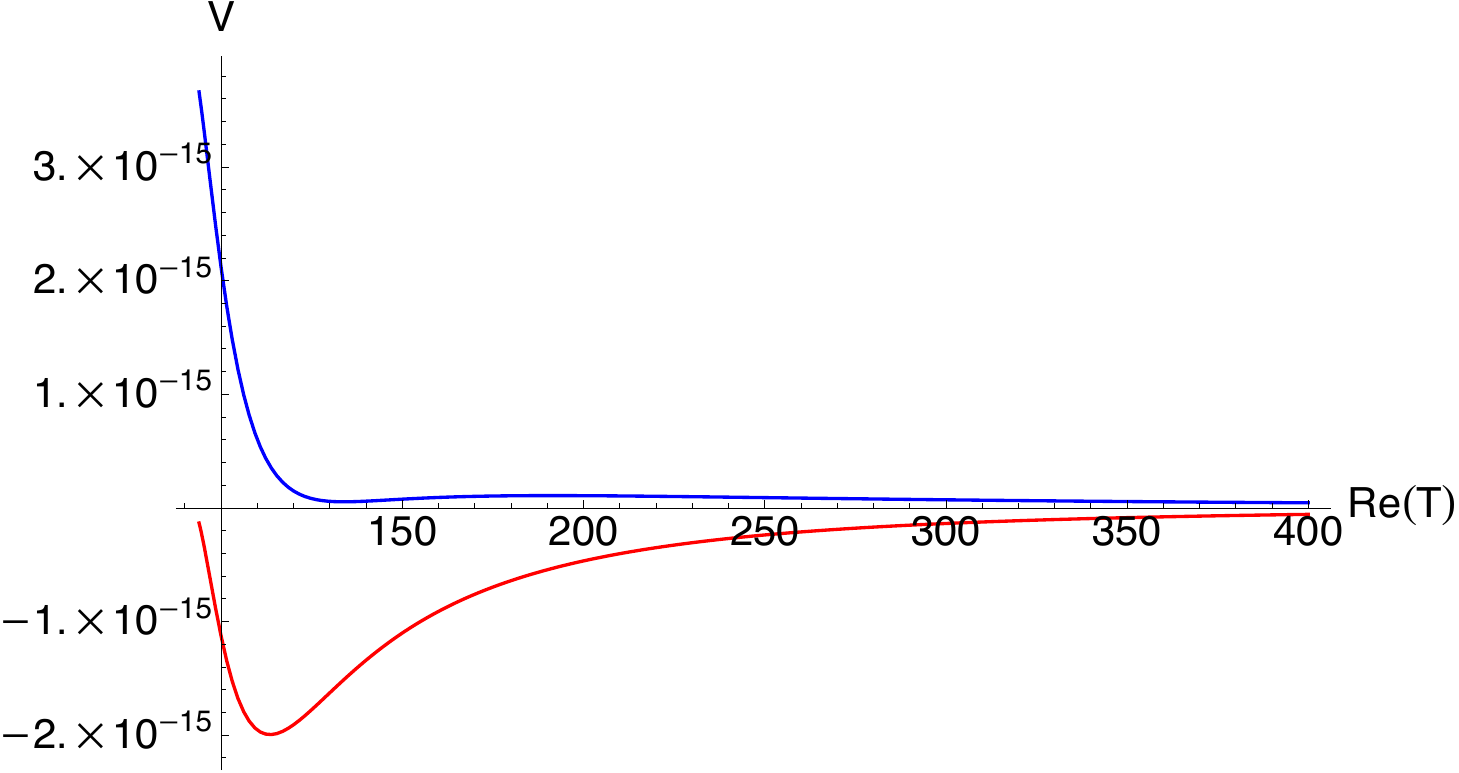} 
   \caption{dS vacuum uplifted from an AdS vacuum by adding a non-perturbative instantonic contribution, with a$=$0.1, A$=$1.0,  $\mathcal{W}_0= 10^{-4}$ and A$_{\widehat{D5}}=3.25 \times 10^{-10}$.}
   \label{fig:example1}
\end{figure}

Although is not possible to analytically solve the equations and find a  critical point for the scalar potential, its existence can be numerically shown.  This minimum is our starting point to construct a controlled dS by incorporating  a non-BPS $\widehat{D5}$ brane which contributes to the scalar potential as
\eq{
V= V_{\text{AdS}} + A_{\widehat{D5}} s^{-1/2} \tau^{-5/2} \,.
}
Thus, as expected, the contribution of this energy to the scalar potential lifts the AdS vacua breaking supersymmetry in a controlled manner parametrized by A$_{\widehat{D5}}$. A stable dS vacuum is possible only for a small value of the pre-factor A$_{\widehat{D5}}$. In Figure \ref{fig:example1}, we present an explicit realization of this mechanism where a the value for $A_{\widehat{D5}}$ was selected by hand.\\

Since the value of $A_{\widehat{D5}}$ is related to the induced metric of the 2-cycle where it is wrapped on, one way to achieve a small value consists of having the non-BPS state located very close to the singular point where the orientifold plane is sitting at the internal tori, such that the volume along the coordinates of the 2-cycle minimizes the tension of the brane. Further studies are required to describe the conditions under which this could be posible\footnote{Notice that in the KKLT scenario, a very similar situation is expected by the inclusion of anti-$D3$-branes where the small value of the pre-factor related to the $\overline{D3}$  depends on both the number of $\overline{D3}$ as well as on the warp factor at the end of the throat.  This is, as the $\overline{D3}$ is gravitationally attracted to the bottom of the throat its energetic contribution is redshifted justifying such a small value required to stabilize the vacuum at a metastable dS one \cite{Aparicio:2015psl}.}. 

\subsection{Model 3: dS by uplifting an asymptotic Minkowski}

In a toroidal model, the condition $\mathcal{W}_{\text{flux}} \ll 1$ is hard to achieve in an explicit case, mainly because the flux component of the superpotential is a polynomial in $U$.  One way to circumvent this issue is to consider a CY compactification with $h^{2,1} = 0$, in the lines of \cite{Blumenhagen:2015kja}. In this case, as we shall shortly show, the moduli can be fixed at a dS vacuum by the inclusion of a non-BPS state which uplifts a runaway direction in a controlled manner.\\

Let us then consider a CY compactification without complex structure, which is similar to an isotropic torus compactification with stabilized complex structure. 
By turning  NS-NS $h$ and RR $f$ fluxes supported on the appropriate cycles, the superpotential given by Eq. (\ref{eq:GVW}) can be written as
\eq{
\mathcal{W}_{\text{flux}} = i f+i S h \,,
}
where the real part of the complex moduli corresponds to the saxionic scalar fields as shown in Eq. (\ref{eq:mod}).  Our strategy here is to analyze regions for large volume such that there is an apparent stable value for $\tau$ with $V\sim 0$. This means that for some specific conditions $\tau$ does not change significantly while $V$ approaches asymptotically to a null value.   In such conditions, we shall show that an uplift to a dS vacuum can be gathered by adding the non-BPS contribution. Let us show the involved details.\\

Consider the superpotential $\mathcal{W}_{\text{inst}}$ in Eq. (\ref{inst}). Since the pre-factor $A$ is a constant, we can rewrite it as $A=\widetilde{A}e^{\alpha}$, with $\alpha$ a  parameter to be used to control the expansion. 
In the limit of large $\tau$, large $\alpha$ and $\alpha-\lambda\tau<1$, the superpotential $\mathcal{W}_{\text{inst}}$ can be approximated, up to third order, to\footnote{An approximation less than third order shows no minimum.} 
\eq{
\mathcal{W}_{\text{inst}} &\approx \widetilde{A} \left( 1+(-\lambda \tau + \alpha)+\frac{(-\lambda \tau + \alpha)^2}{2} + \frac{(-\lambda \tau + \alpha)^3}{3!} \right) e^{\text{i}\lambda\rho}\,.
}
Let us now proceed to stabilize the moduli $S$ and $T$ by considering the whole superpotential  $\mathcal{W}=\mathcal{W}_{\text{flux}}+ \mathcal{W}_{\text{inst}}$ where we take the approximate expression for $\mathcal{W}_{\text{inst}}$. By vanishing the K\"ahler derivative $D_S \mathcal{W}=0$, the axio-dilaton is fixed at
\begin{equation}
S_0=\frac{\bar{\mathcal{W}}_{\text{inst}}(\tau_0)-if}{ih},
\end{equation}
where $\tau_0$ is the value for $\tau$ for which $\partial_T V=0$. The K\"ahler modulus is then approximately fixed at

\eq{ \label{eq:vevsads}
\tau_0 &\approx \frac{\gamma+\alpha}{\lambda} \, \quad \rho_0  =0\,, 
}
where 
\begin{equation}
\gamma =1-\frac{1}{(1+\sqrt{2})^{1/3}}+(1+\sqrt{2})^{1/3}\approx 1.59607.
\end{equation}
Then, $\mathcal{W}_{\text{inst}}(\tau_0)\approx\widetilde{A}e^{-\gamma}$ and we can safely consider that for sufficiently large $\tau_0$ (i.e. for large $\alpha$ and in consequence a small $\widetilde{A}$) $\mathcal{W}_{\text{inst}}(\tau_0)\approx 0$. It is important to emphasize that $V$ has not a true minimum along $\tau$, but for large volume, it is possible to find a region in which $\tau$ changes very slowly while $V$ is given by
\begin{equation}
V(\tau_0,s_0)\approx \frac{4}{3}A^2\lambda^2\tau_0^2e^{-2\lambda\tau_0},
\end{equation}
which approaches zero for large $\tau_0$. Since $V$ approaches zero from positive values, we are in a non-supersymmetric region where
\eq{
D_T \mathcal{W} \approx -\lambda \widetilde{A} e^{-\gamma}-\frac{3if}{\tau_0}\approx -\frac{3if}{\tau_0},
}
and where we have used that
\begin{equation}
s_0\approx \frac{f}{h}, \qquad c_0\approx 0.
\end{equation}

%

Thus, we observe that SUSY is broken in a controlled manner parametrized by $\tau_0 \gg 1$, For large $\alpha$ and for positive $f > h$, this region provides physical values for the moduli while controlling corrections to the K\"ahler potential.  The asymptotically zero value for $V$ can be understood as a cancellation between the F-term contribution to the scalar potential which main contribution scales as $D^2 \mathcal{W} \sim f^2$ and the contribution of $\mathcal{W}_{\text{flux}} \sim f^2$.  As expected in a runaway direction, we can also see that for large $\tau_0$ the mass eigenvalues tends to vanish. In fact,
\eq{
m^2 \approx \left(  \frac{A^2 e^{-2 \alpha} \lambda^5 h}{2^5 (\alpha+\gamma)^5 f} p(\alpha),  \frac{h^3 \lambda^3 }{2^3 f (\alpha +\gamma )^3}   \right) \,.
}
approaches to zero for large $\alpha$ (or equivalently for large $\tau_0$), where $p(\alpha)$ is a polynomial on $\alpha$. The question here is if it is possible to uplift this potential to a stable dS vacuum. The key point is to realize that the scalar potential is exponentially suppressed, for which an uplift depending on a power of $\tau$ could give us a stable vacuum. Let us then consider the presence of a 
 non-BPS state, meaning that we have the scalar potential
 \begin{equation}
 V_{\text{total}}=V+ V_{\widehat{D5}}.
 \end{equation}
 
It is hard to construct an analytic expression for the vacua, thus we proceed numerically. We find that stable dS minimum can be constructed by uplifting regions of large $\tau_0$ where  the v.e.v. of the saxionic components are shifted to higher values and a metastable dS is found at
\eq{
\rho = \frac{\pi}{2 \lambda} \quad c = 0.
}

\begin{figure}[htbp]
   \centering
  a) \includegraphics[scale=0.28]{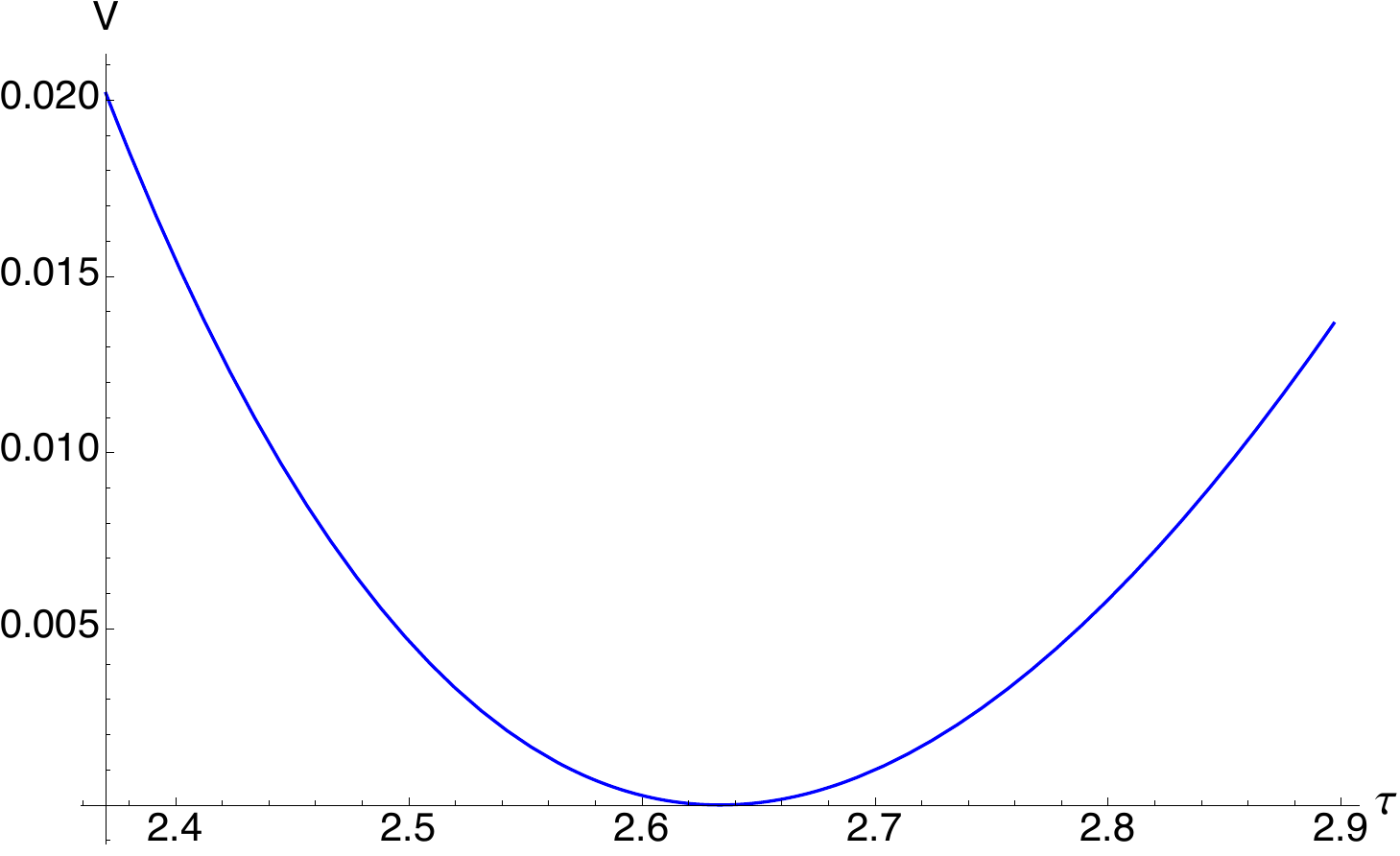} b)  \includegraphics[scale=0.28]{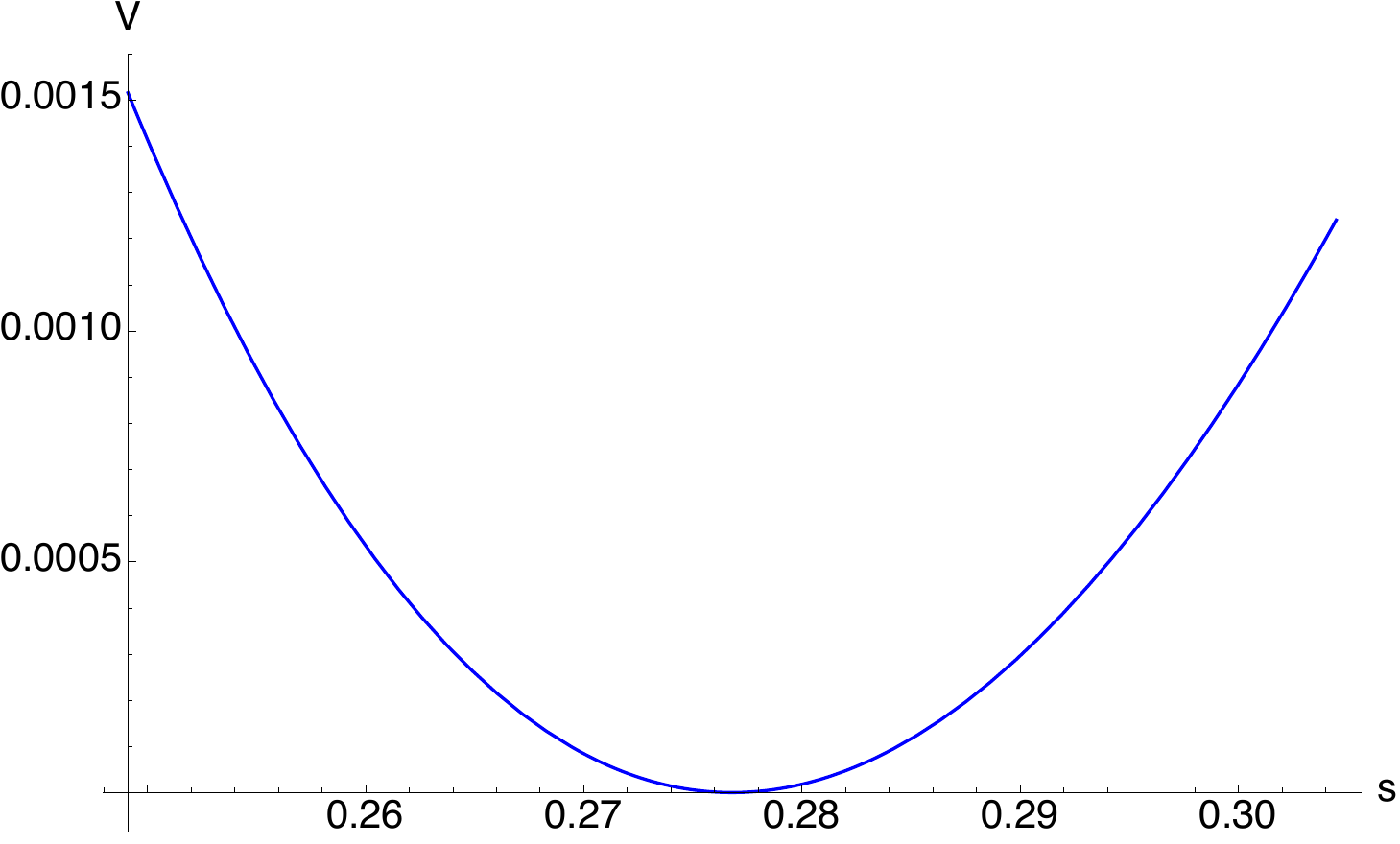} 
   \caption{dS vacua constructed from a scalar potential with non-BPS states, taking $f=$2, $h=$-12,  A$_{\widehat{D5}}= 2 \times 10^{-24}$, A$=-1.8923$ and $\lambda = -2.1735$. Figure a( shows the minimum in $\tau$ while figure b) shows it for $s$.}
   \label{fig:example2}
\end{figure}

In Figure \ref{fig:example2} we present a particular case where a metastable dS vacuum is found. Remember that we are working on an approximate value for the non-perturbative instantonic superpotential for large values of $\tau$.\\

 As mentioned above, the small value required for $A_{\widehat{D5}}$ can be justified if the non-BPS brane reduces its volume along the 2-cycle around the orientifold fixed point. In the case of a CY this can also be justified if
the orientifold planes are located at a warped throat where strong gravitational effects redshift its energetic contribution. However, any back reaction in the geometry of the internal space may destabilize the vacuum, and its explicit stability shall be addressed carefully. As in the previous case, more studies are required.\\

\section{Instabilities from brane-flux transitions}

As we have shown, by considering a single non-BPS $\widehat{D5}$-brane, it is possible to construct a dS minimum under some specific conditions. In this section, we want to show that even if we do not consider the cancellation of the total discrete K-theory charge, the involved process due to the existence of 3-form fluxes, induces some brane-flux transitions that could source for instabilities of the afore-constructed dS minima in agreement with the RdSC. At the end, this will leads us to a new perspective of how to cancel the discrete K-theory charge while keeping the possibility of constructing a dS vacuum, which nevertheless, makes it unstable.\\

Let us start by applying the AHSS described in section 2 to the non-BPS $\widehat{D5}$-brane. As mentioned, the first step in the AHSS must involve mappings from the torsional components of the cohomology groups,
\eq{
 d_3: \text{Tor}\,\H^p(T^6/\mathbb{Z}_2;\mathbb{Z})\longrightarrow \text{Tor}\,\H^{p+3}(T^6/\mathbb{Z}_2;\mathbb{Z}),
 }
 through the NS-NS differential $d_3$. The possible mappings are shown in the diagram of Figure \ref{fig:mappings}.\\

\begin{figure}[!htbp]
   \centering
  
  \begin{tikzpicture}[
roundnode/.style={circle, draw=orange!80, fill=white!10, very thick, minimum size=7mm},
squarednode/.style={rectangle, draw=red!80, fill=white!5, very thick, minimum size=5mm},
]
\node[]  at (0,0)		(1)     {Tor H$^0(T^6/\mathbb{Z}_2;\mathbb{Z})\qquad \qquad\qquad$};
\node[]  at (0,1.0) 	(2)      {Tor H$^1(T^6/\mathbb{Z}_2;\mathbb{Z})\qquad \qquad\qquad$};
\node[]     at (0,2.0) (3)      {Tor H$^2(T^6/\mathbb{Z}_2;\mathbb{Z})\qquad \qquad\qquad$};
\node[]    at (0,3.0)  (4)      {Tor H$^3(T^6/\mathbb{Z}_2;\mathbb{Z})\qquad \qquad\qquad$};
\node[]     at (0,4.0)  (5)       {Tor H$^4(T^6/\mathbb{Z}_2;\mathbb{Z})\qquad \qquad\qquad$};
\node[]     at (0,5.0)  (6)        {Tor H$^5(T^6/\mathbb{Z}_2;\mathbb{Z})\qquad \qquad\qquad$};
\node[]     at (0,6.0)  (7)       {Tor H$^6(T^6/\mathbb{Z}_2;\mathbb{Z})\qquad \qquad\qquad$};
\node[]    at (3,6.0)   (7d)       {$\qquad\qquad \qquad \text{Tor}\,\H^6(T^6/\mathbb{Z}_2;\mathbb{Z})$};
\node[]    at (3,5.0)   (6d)     {$\qquad\qquad \qquad \text{Tor}\,\H^5(T^6/\mathbb{Z}_2;\mathbb{Z})$};
\node[]    at (3,4.0)   (5d)      {$\qquad\qquad \qquad \text{Tor}\,\H^4(T^6/\mathbb{Z}_2;\mathbb{Z})$};
\node[]    at (3,3.0)   (4d)      {$\qquad\qquad \qquad \text{Tor}\,\H^3(T^6/\mathbb{Z}_2;\mathbb{Z})$};
\node[above]    at (1.5,1.7)   (d3)      {d$_3$};

\draw[arrows = {-Stealth[length=10pt, inset=2pt]}] (1) -- (4d);
\draw[arrows = {-Stealth[length=10pt, inset=2pt]}] (2) -- (5d);
\draw[arrows = {-Stealth[length=10pt, inset=2pt]}] (3) -- (6d);
\draw[arrows = {-Stealth[length=10pt, inset=2pt]}] (4) -- (7d);

\end{tikzpicture}

     \caption{Map of d$_3$ operator. We are interested in the mappings between the torsional components of the cohomology groups.}
   \label{fig:mappings}
\end{figure}

In our case, a 4-form $\omega_4 \in \text{Tor}\,\H^4(T^6/\mathbb{Z}_2;\mathbb{Z})$ represents the non-BPS $\widehat{D5}$. Since $d_3(\omega_4)=0$, this brane is stable unless it encounters an instantonic brane. The existence of an instantonic brane can be deduced from the mapping
\begin{equation}
d_3: \text{Tor}\,\H^1(T^6/\Z_2;\mathbb{Z})\rightarrow \text{Tor}\,\H^4(T^6/\Z_2;\mathbb{Z}),
\end{equation}
given by
\begin{equation}
d_3(f_1)= H_3\wedge f_1=\omega_4,
\end{equation}
with $f_1\in \text{Tor}\,\H^1(T^6/\mathbb{Z}_2;\mathbb{Z})$. Under Poincar\'e duality a 1-form is related to a torsional 5-cycle in $T^6/\mathbb{Z}_2$. This corresponds to a $D7$-brane wrapping an internal discrete 5-cycle (which in a toroidal compactification is easy to construct)
in $\text{Tor}\,\H_5(T^6/\mathbb{Z}_2;\Z)$ and the extended 3 spatial coordinates. This is an instantonic 7-brane\footnote{Actually it is an anti D7-brane \cite{Loaiza-Brito:2004ajy}.} which drives a transition between non-BPS branes and torsional fluxes.\\


The brane-flux transition can be represented as follows: the non-BPS $\widehat{D5}$-brane encounters an instantonic $D7$-brane  supporting the NS-NS $H_3$ flux and it transforms into a discrete (half-integer modulo 2) flux $f_1$, such that the discrete topological charge carried by the non-BPS $\widehat{D5}$-branes is now carried by the fluxes $H_3$ and $f_1$, this is
\begin{equation}
\int_{\Sigma_4} H_3 \wedge f_1 = 1 \, \text{mod} \, 2 \,,
\end{equation} 
where $\Sigma_4$ is a torsional 4-cycle in the internal space, transversal to the 2-cycle where the original non-BPS $\widehat{D5}$-brane was wrapped.\\

In this context a $\widehat{D5}$-branes is equivalent to a flux configuration given by the NS-NS 3-form flux $H_3$ and a discrete 1-form flux $f_1$ belonging to $\text{Tor}\,\H^1(T^6/\Z_2) = \Z_2$ (i.e. that $2 f_1 = d\sigma_0$ for a 0-form $\sigma_0 \in \H^0(T_6/\Z_2)$). \\


Hence, since the original configuration consisting of a $\widehat{D5}$-brane and 3-form fluxes $H_3$ has transformed into a configuration with a discrete $f_1$-form and the same NS-NS 3-form fluxes, the contribution of the non-BPS state in the scalar potential changes as\footnote{The moduli dependence of the discrete flux $f_1$ can be derived directly from dimensional reduction of the RR flux term \cite{Hertzberg:2007wc}.}
\begin{equation}
V_{\hat{D5}}\sim \frac{A_{\widehat{D5}}}{{s^{1/2}\tau^{5/2}}} \longrightarrow V_{f_1}\sim \frac{A_{f_1}}{s^2\tau^2},
\end{equation}
thus, the minimum before the transition suffers a transformation (if the internal geometry admits the presence of torsional 1-forms). This means that the aforementioned de Sitter (dS) vacuum has become destabilized. It is important to note that this topological transition can continue if it is energetically favorable. The different vacua that are connected through instantonic branes could be seen as a mechanism to modify seemingly stable vacua\footnote{It is interesting to observe that the new configuration is dependent on torsional fluxes $f_1$, which can be interpreted as fractional, as explained in \cite{CaboBizet:2019sku,Damian:2018tlf}.}.\\

Finally, it should be noted that the T-dual version of the  $\widehat{D5}$-brane is represented by a non-BPS $\widehat{D6}$-brane in type IIA which wraps a 3-cycle that could support some units of NS-NS flux, rendering the state to be FW anomalous. This $\widehat{D6}$-brane is represented by a non-closed 3-form under $d_3$, i.e., $d_3\omega_3\ne 0$. Under T-duality, we can say that a $d_3$ non-closed form transforms into a $d_3$ exact form since the $\widehat{D5}$ in type IIB is represented by a 4-form such that $\omega_4=d_3 f_1$.\\

\subsection{A proposal for canceling discrete K-theory}

Tadpole cancellation in conventional flux compactifications can be reinterpreted in terms of flux-brane transitions that are derived from the MMS instantons and the AHSS. To see this, let us consider $n$ D3-branes located at a point in $X_6$, and $m$ units of NS-NS flux $H_3$ supported on a 3-cycle of $X_6$. Since the D3-branes are represented by a 6-form $\omega_6$, it is clear that $d\omega_6=0$ and $d_3\omega_6=0$. However, there exists a non-trivial map $d_3: \H^3(X_6;\mathbb{Z})\rightarrow H^6(X_6;\mathbb{Z})$ that indicates the presence of an instantonic $D5$-brane, which is represented by a 3-form $F_3$ with $d_3(F_3)=\omega_6$. The instantonic brane drives the transition of $m$ D3-branes into a unit of RR flux $F_3$. As a result, the charge of the disappeared D3-branes is now carried by the fluxes through the coupling $F_3\wedge H_3$, and the total charge is conserved after the transition. This means that the quantity
\eq{
N_{D3}+\int_{X_6}F_3\wedge H_3,
}
remains constant under these topological transformations. As the orientifolds provide a negative amount of charge, the total D3-brane charge must vanish.\\

The same argument can be applied to our discrete case, leading to the conclusion that
\eq{
N_{\widehat{D5}}+\int_{T^6/\mathbb{Z}_2} f_1\wedge H_3 = 0 \,\text{mod} \,2.
\label{dtadpole}
}

Hence, although it is possible to cancel all discrete K-theory by considering an even number of non-BPS states, it is also possible to cancel it by considering a single non-BPS state (or equivalently an odd number) and a discrete flux. Since they have a different dependence on moduli, it is possible to construct a dS extreme solution even with the total discrete K-theory charge equal to zero. However, as the transition occurs, it is enough for one of them to change into the other for the total charge to explicitly cancel, destroying any possible vacuum and rendering the dS vacua to lie in the Swampland.\\

An interesting question arises regarding the time-average of these processes, with the intention of determining if they are allowed under the TCC. We expect a short-lived scenario since an instantonic 7-brane is a very heavy soliton solution. A much more detailed study is required to answer this question, which we leave for future work.\\

\section{Final Comments}

We discuss some general issues regarding the construction and stability of a de Sitter (dS) vacuum in toroidal compactifications with 3-form fluxes and orientifold three-planes. We show that it is possible to construct an apparently stable dS vacuum by incorporating a non-BPS $\widehat{D5}$-brane and uplifting either an asymptotically Minkowski scalar potential or an Anti-de Sitter (AdS) vacuum. However, the uplifted potential becomes very flat, indicating that the vacuum may be easily perturbed by a short-lived dS minimum, placing such scenarios in the Landscape, in agreement with the Trans-Planckian Censorship Conjecture. Nevertheless, these vacua cannot be consistent if the K-theory discrete charge, carried by the non-BPS brane, must cancel in compact spaces as expected by the Cobordism Conjecture, implying that any consistent scenario in the Landscape must contain an even number of $\widehat{D5}$-branes.\\

We show that even if we ignore the cancelation of discrete K-theory charge (for a while), the presence of NS-NS fluxes necessary to stabilize some of the moduli makes the non-BPS $\widehat{D5}$-brane unstable to transform into discrete flux $f_1$, which now carries a $\Z_2$-charge upon the appearance of an instantonic D$7$-brane. This topological instability of the non-BPS $\widehat{D5}$-brane is the T-dual manifestation of the Freed-Witten anomaly present in a scenario where a non-BPS $\widehat{D6}$-brane in type IIA wraps a 3-cycle supporting a NS-NS flux. It is worth noting that the transition only affects the $\widehat{D5}$-brane required to construct the dS vacuum. In the absence of non-BPS states, the vacuum is AdS, and any topological transformation between 3-form RR fluxes and $D3$-branes does not change the scalar potential, making these vacua stable under such transitions.\\

 Since the scalar potential's dependence on the moduli is different for the discrete 1-form and the $\widehat{D5}$-brane, it is not certain that the new configuration will have the same minimum, and the seemingly stable dS vacuum could become destabilized. Furthermore, since the NS-NS flux remains, a sequence of transformations may occur, but the chain would end when a transition requires more energy than the previous step.\\

Therefore, just as with integer charges, having different sources for a discrete K-theory charge allows cancelation while having a non-trivial contribution to the scalar potential. Cancelation of the discrete charge requires a single (or odd number of) $\widehat{D5}$ and a 1-form flux $f_1$ such that Eq. (\ref{dtadpole}) is satisfied. This allows scenarios where the total discrete K-theory charge vanishes while having a contribution of $\widehat{D5}$ and $f_1$ to the scalar potential, upon which it is possible to construct dS vacua since their contributions to the scalar potential have  different dependence on the K\"ahler modulus. \\

However, these vacua can suffer two instabilities: 1) if the potential is very flat, they may be very short-lived, and ii) due to topological transitions and depending on the internal geometry, where non-BPS states wrap 2-cycles and discrete fluxes are supported on 1-cycles, their contribution to the scalar potential can change, thereby destroying the original vacuum. Considering all these processes, we can conclude that the dS vacua constructed in a scenario with zero discrete charge, are in the Landscape according to the Trans-Planckian Censorship, the Refined de Sitter, and the Cobordism conjectures. We leave the study of more general scenarios and a more detailed verification of the Swampland conjectures for future work.

\vspace{1cm}

\begin{center}
{\bf Acknowledgments}
\end{center}
C.D. is supported by CIIC-UG-DAIP No. 126/2022 while  O. L-B is supported by CONACyT project CB-2015, No. 258982 and by CIIC-UG-DAIP No. 236/2022.

\bibliographystyle{JHEP}
\bibliography{References}

\end{document}